\documentclass{ifacconf}
\usepackage{bm}                 
\usepackage{amsmath}
\usepackage{booktabs}
\usepackage{array}
\usepackage{multirow}
\usepackage[utf8]{inputenc}
\usepackage{float}
\usepackage{graphicx}  
\usepackage{cellspace}
\usepackage{makecell}
\usepackage{natbib}        

\begin{document}
\begin{frontmatter}

\title{Efficient Safety Verification of Autonomous Vehicles with Neural Network Operator\thanksref{footnoteinfo}} 

\thanks[footnoteinfo]{This work was supported by National Natural Science Foundation of China No.62473224, Beijing Natural Science Foundation 4244092, and sponsored by Beijing Nova Program 20230484259 and 20240484642. \textit{(Corresponding author: Shuo Feng)}}

\author[First]{Lingxiang Fan} 
\author[Second]{Linxuan He} 
\author[Third]{Haoyuan Ji}
\author[Correspond]{Shuo Feng}
\address[First]{Department of Automation, Tsinghua  University, Beijing 100084, China (e-mail:fanlx23@mails.tsinghua.edu.cn)}
\address[Second]{Department of Automation, Tsinghua  University, Beijing 100084, China (e-mail:hlx24@mails.tsinghua.edu.cn)}
\address[Third]{Department of Automation, Tsinghua  University, Beijing 100084, China (e-mail:jihy21@mails.tsinghua.edu.cn)}
\address[Correspond]{Department of Automation, Tsinghua University,
       Beijing National Research Center for Information Science and Technology (BNRist),  Beijing 100084, China (e-mail:fshuo@tsinghua.edu.cn)}

\begin{abstract}                
When autonomous vehicles encounter untrained scenarios, ensuring safety hinges on effective safety verification to prevent accidents stemming from unexpected model decisions. Reachability analysis, a method of safety verification, offers relatively high precision but at the cost of significant computational complexity. Our method leverages end-to-end neural network operators to compute reachable sets, replacing traditional mathematical set operators, thereby achieving higher efficiency in safety verification without substantially compromising accuracy or increasing conservativeness. We define vehicle dynamics on discrete time series and detail the safety verification process and safety standard based on reachable sets. Experimental evaluations conducted in several typical road driving scenarios demonstrate the superior efficiency performance of our proposed operator over classical methods.
\end{abstract}


\end{frontmatter}

\section{Introduction}
\label{sec1}

With the rapid advancement of artificial intelligence, the performance of autonomous driving models has significantly improved, making autonomous driving technology a viable and implementable solution. For instance, the autonomous driving model \citep{c1}, developed using large-scale models, demonstrates high performance in tasks such as perception and prediction, effectively addressing the complexity of driving scenarios and the diversity of tasks. By training on data collected from real-world or simulated driving environments, these models can learn robust strategies to handle scenarios encountered during training. However, when faced with untrained scenarios in real-world applications, the model's strategies and outcomes may become unpredictable, potentially leading to severe safety incidents. This unpredictability poses a significant barrier to the large-scale commercial deployment of high-level autonomous vehicles \citep{c2}.

To address these challenges, safety verification \citep{c3} is essential. This involves using mathematical computations to prove that autonomous vehicles will not cause safety accidents due to uncertainties in real-time decision-making. Such verification ensures the safety of decisions made by autonomous vehicles.

One prominent method for safety verification is reachability analysis. This approach represents the physical state of a vehicle using mathematical sets, evolves these sets based on the vehicle's dynamics to deduce reachable sets, and then verifies safety based on these sets \citep{c4}. The evolution process can be abstracted as mathematical operations, such as linear transformations and summations of sets, ensuring both accuracy and feasibility of the decisions made by autonomous vehicle.

Reachability analysis can be categorized into two types. The first type involves precise element-by-element set operations \citep{c5}, which produce accurate results but are computationally intensive. Currently, it has been successfully applied to simple autonomous driving scenarios. For example, a study calculates the reachable sets of vehicles in a scenario where two autonomous vehicles cut into each other's paths, verifying whether these sets intersect to avoid collisions \citep{c6}. Another study focuses on a seven-dimensional vehicle system, providing additional safety control strategies based on the application of reachable sets \citep{c8}. However, in high-dimensional real-world driving scenarios, the computational complexity of set operations becomes prohibitive, resulting in significant computational cost. Given the short decision-making windows in autonomous driving, real-time online safety verification in complex environments remains unfeasible.

The second type of reachability analysis improves efficiency by sacrificing some accuracy to reduce computational complexity. This is achieved by merging and simplifying the elements of the set and performing approximate set operations to deduce reachable sets. While this approach introduces some errors, it offers higher efficiency and scalability. Recent research has explored calculating approximate reachable sets using predivided base sets, enabling real-time route re-planning based on the safety of these sets \citep{c9}. Other studies analyze collision risks by checking whether the approximate reachable sets of vehicles avoid obstacles, employing vehicle coordination or other strategies to ensure safety through online analysis of these sets \citep{c10,c12, c13}. In addition, some research focuses on assessing collision risk by determining whether the approximate reachable sets of vehicles can avoid obstacles \citep{c14} or calculating the necessary time buffers to ensure safety. However, because this type of reachability analysis relies on conservative assumptions, it often overestimates risks, leading to overly cautious driving strategies. This conservativeness can reduce driving efficiency and render the feasible region infeasible.

Given the limitations in these two approaches, we propose using neural network operator \citep{c31} to perform efficient set operations based on approximate reachability analysis. By training the neural network operator on accurate cases generated by comprehensive reachability analysis algorithms in specific scenarios, we can transform the highly complex set operation process into a more efficient alternative without significantly compromising accuracy or introducing excessive conservativeness.

Several related works have investigated the application of neural networks to reachability analysis in hybrid systems. Phan et al. \citep{c32} employ neural networks to address the state classification problem for hybrid systems, while Rubies-Royo et al. \citep{c33} introduce a neural network classifier to approximate the optimal controller for control-affine systems. However, their networks do not end-to-end learn the mapping from the current state to the future reachable set; instead, they address specific sub-problems, and their outputs must be integrated with other analytical methods to eventually deduce the reachable set and provide a safety assessment, making their method less efficient. Sergiy Bogomolov et al. \citep{c31} also propose the use of neural networks to accelerate the reachability analysis process in safety verification. However, their method fails to reduce the dimensionality of the reachable set, resulting in excessively large data dimensions and challenges in neural network training. As a result, their approach is ineffective in complex vehicle driving scenarios. 

In this work, we propose a novel safety verification framework that provides end-to-end reachability analysis-based safety verification. We employ neural network operator in reachable set deduction to replace the traditional mathematical set operators, thereby reaching a higher efficiency. Unlike other neural network-based methods, we incorporate more precise vehicle dynamics and dimensionality-reduced set representation in the process of generating training data, which enables the operator to address reachable sets from scalable vehicle systems. Experimental evaluations conducted in several
typical road driving scenarios demonstrate the superior efficiency performance of our proposed operator over traditional
methods. Overall, our primary contribution lies in a holistic approach that addresses the key limitations of prior reachability analysis methods, specifically:

\begin{itemize}
    \item \textbf{End-to-end neural network operator for efficient reachable set deduction.}  Our method proposes an end-to-end neural network operator that directly deduces the  reachable set based on the input state, replacing the computationally expensive core of traditional reachability analysis and providing efficient online safety verification. 
    \item \textbf{Dimensionality reduction in reachable set representation.} We introduce a Zonotope-based reachable set representation that reduces the feature dimension from the order of $10^5$ to $10^1$. It enables the reachable set deduction to perform effectively in complex vehicle systems.

    \item \textbf{Enhanced Fidelity with Dynamic Models.} We advance beyond the simple kinematic models used in other neural network-based methods, providing more precise vehicle dynamics. This vehicle dynamics is employed in the process of generating training data for the neural network operator. As a result, it enables more accurate and physically plausible safety verification. 
\end{itemize}

The remainder of this paper is organized as follows: In Section \ref{sec2}, we present the notation used throughout this work, the set representation of vehicle states, the safety standards for vehicles, and the neural network we employed. Section \ref{sec3} provides a formal problem statement. In Section \ref{sec4}, we describe our proposed methodology, including the vehicle dynamics model, the data generation process, the detailed structure of the neural network and the safety targets for the model. Section \ref{sec5} presents the results of model training and evaluates the model's performance across various scenarios. Finally, Section \ref{sec6} concludes the paper.

\section{Preliminary}
\label{sec2}
\subsection{Notation}

We begin by defining the vehicle system to be verified on a discrete time series. The position, speed, direction, and other physical quantities of the vehicle at time $t$ are represented by a state vector, defined as $x(t)\in R^n$. At time $t$, the vehicle receives control signals for physical quantities such as acceleration and orientation, defined as $u(t)\in R^m$. Let $\Delta t$ denote a short time interval. In the ideal case, the state of the vehicle at time $t+\Delta t$ can be calculated from $x(t)$ and $u(t)$, as defined by:
\begin{equation}
x(t+\Delta t)=x(t)+\Delta x(t)=x(t)+f(x(t),u(t))\cdot \Delta t.\label{eq1}
\end{equation}

Here, $f$, referred to as the dynamic function, is a specific continuous function for a given vehicle. The process of deducing $x(t+\Delta t)$ from $x(t)$ is termed one step, and $N\in Z^+$ is defined as the preset number of safety verification steps. The safety verification of autonomous driving involves deducing the states of autonomous vehicles and other background vehicles from the current time, given the control signal at the current time. It then checks the safety of the states $x(t+\Delta t),x(t+2\Delta t),...,x(t+N\Delta t)$, and determines the feasibility of the control signal based on the safety of these states.

\begin{figure}[thpb]
    \centering
    \includegraphics[width=2.5in]{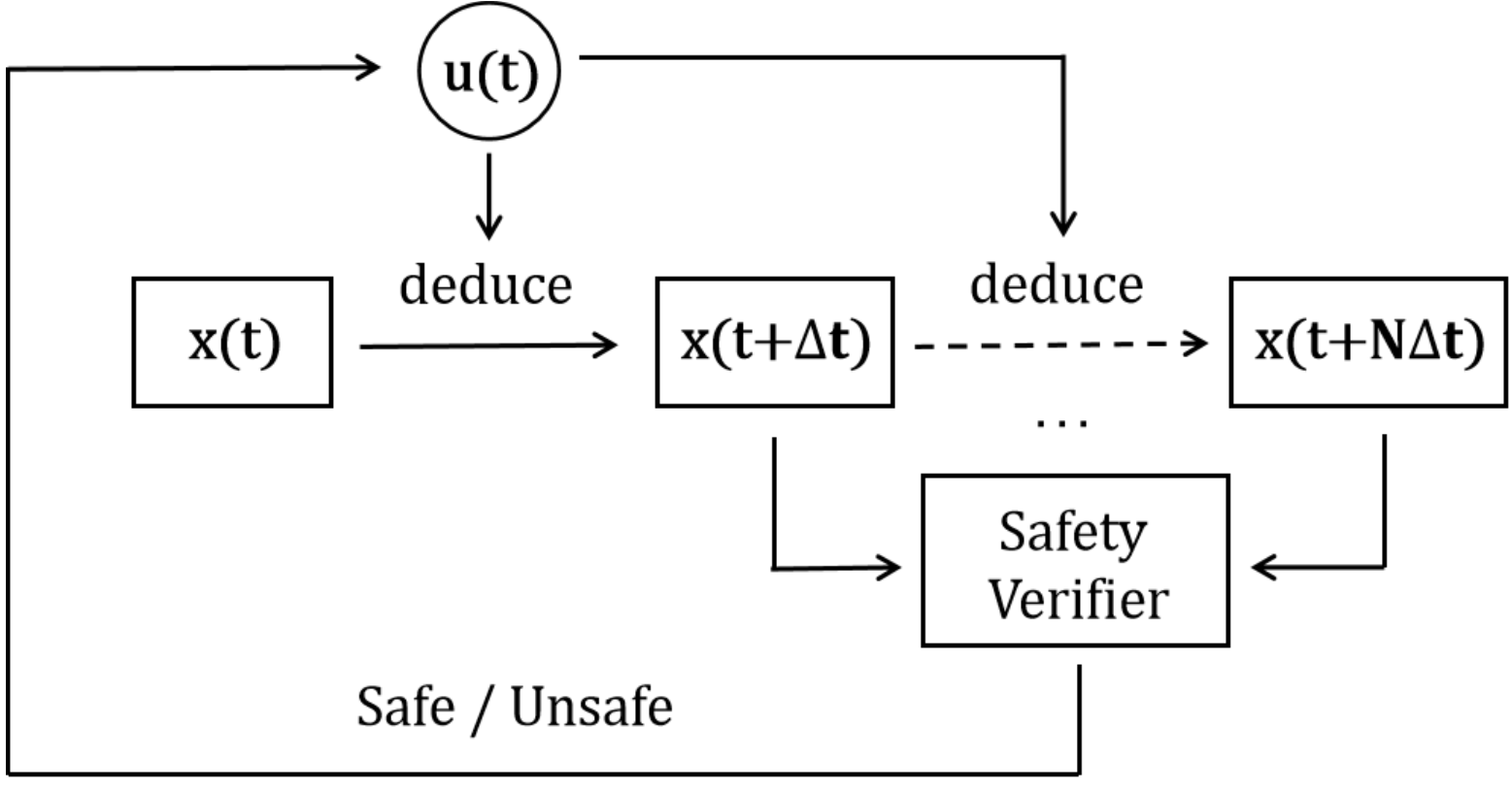} 
    \caption{An overview of safety verification process} 
    \label{fig1}
\end{figure}

In real-world traffic scenarios, autonomous vehicles face uncertainties in real-time decisions made by other background vehicles and errors in the observation of vehicle states. To account for these uncertainties, we characterize each physical quantity of the vehicles as a bounded variable, defined by an upper bound ($ub$) and a lower bound ($lb$). Specifically, for any $x_i$ in $x=\left(x_1,x_2,...,x_n\right)$, $x_i=[lb_i,ub_i]$, the physical quantity represented by $x_i$ can take any value within the range $[lb_i,ub_i]$. Consequently, $x(t+\Delta t)$ cannot be directly deduced from $x(t)$ and $u(t)$ using Eq. \eqref{eq1}. Instead, set-based reachability analysis is employed to address this challenge.

We employ mathematical sets of high-dimensional vectors to represent all possible values that the vehicle states or control signals can assume. The set containing all the possible vehicle states at time $t$ is defined as $X(t) \subseteq R^n$. The control signal set at time $t$ is defined as $U(t) \subseteq R^m$. 

\subsection{Safety Standard}

Consider the sets $X_0(t),X_1(t),...,X_l(t)\subseteq R^n$, where each set $X_i(t)$ represents the reachable state of a vehicle $A_i$ at time $t$. Specifically, $X_0(t)$ corresponds to the autonomous vehicle under verification, while the remaining sets describe the reachable states of the background vehicles. The system is deemed safe at time $t$ if and only if both predefined safety standards are satisfied.

\subsubsection{Safety for each vehicle(S1)}

Given safety constraint matrices $C_{s0},C_{s1},...,C_{sn}\in R^{q\times n}$ and safety constraint vectors $d_{s0},d_{s1},...,d_{sn}\in R^q$, define the safe state set of each vehicle $S_0,S_1,...,S_l\in R^n$ as:
\begin{equation}
S_i:=\{x|C_{si}x\le d_{si}\}, \forall i\in [0, l],\label{eq6}
\end{equation}
which are not determined by time.

Set $X_i(t)$ satisfies this safety standard if and only if it satisfies:
\begin{equation}
X_i(t)\subseteq S_i.\label{eq7}
\end{equation}

\subsubsection{Safety between vehicles(S2)}

To verify that there is no risk of collision between the ego vehicle and any background vehicle, we examine the overlap of their physical boundaries. Let $Y_0(t),...,Y_l(t) \subseteq R^2$ denote the sets representing the occupied area of each vehicle in the 2D plane (e.g., the octagons shown in Fig. 4), which are derived from their respective state sets $X_0(t),...,X_l(t)$ by accounting for the vehicle's shape, position, and orientation. The vehicle system satisfies this safety standard if and only if the ego vehicle's occupied area does not intersect with that of any background vehicle:
\begin{equation}
Y_0(t) \cap Y_i(t) = \emptyset, \quad \forall i \in [1, l].
\label{eq8}
\end{equation}

\section{Problem statment}
\label{sec3}
Let $a_0$ denote the autonomous vehicle subject to safety verification, referred to as the ego vehicle, and let $a_1,a_2,...,a_l$ represent background vehicles that may interact with the ego vehicle. Let $A(a_1,a_2,...,a_l,t)$ denote the vehicle system within a specific traffic scenario. The system comprises:

(1) Deduction interval and total step: $\Delta t$ and $N$.

(2) States: $x_0(t),x_1(t),...,x_l(t)\in R^n$. These can be transformed into Zonotopes $X_0(t),X_1(t),...,X_l(t)\subseteq R^n$ using Eq. (\ref{eq3}-\ref{eq4}).

(3) Control signals: 
\label{control}
$u_0(t)\in R^m$. These can be transformed into a Zonotope $U_0(t)\subseteq R^m$ using Eq. (\ref{eq3}-\ref{eq4}).

(4) Dynamic functions: $f_0,f_1,...,f_l:R^n\to R^n$. We assume
all vehicles share the same physical properties, thus using $f$ for all.

(5) Controller: A function to compute $u_0(t+i\Delta t),...,u_l(t+i\Delta t)$ by $x_0(t+i\Delta t),...,x_l(t+i\Delta t)$. We preset maximum increments and decrements $\delta_{+}, \delta_{-}\in R^m$. Thus, $u_0(t+i\Delta t)=[u_0-\delta_{-}, u_0+\delta_{+}],\forall i\in [1,N]$. Additionally, to generalize under uncertain decisions of $a_1,...,a_l$, we preset maximum and minimum control signals $u_+, u_-\in R^m$ for background vehicles. Thus, $u_j(t+i\Delta t)=[u_-, u_+], \forall i\in [0, N], \forall j\in [1, l]$.

(6) Safety Standards: $S1, S2$. Defined by Eq. (\ref{eq6}-\ref{eq8}).

These constitute the inputs to our problem. The boolean variable $flag$ is unknown, indicating whether $u_0(t)$ is safe.

Our work focuses on designing a model $F$ to perform the deduction process illustrated in Figure \ref{fig1} and determine $flag$ based on all inputs. Specifically, $F$ is a neural network operator. The problem statement is listed as in Table \ref{t1}.

\begin{table}[ht]
\centering
\caption{Problem statement}
\renewcommand{\arraystretch}{1.5}
\begin{tabular}{|c|c|c|c|}
\hline
\textbf{System} & \textbf{Inputs} & \textbf{Model} & \textbf{Unknown} \\
\hline
\multirow{5}{*}{$A(a_0, a_1,..., a_l, t)$} 
& $\Delta t$, $N$ & \multirow{5}{*}{$F$} & \multirow{5}{*}{$flag$} \\
\cline{2-2}
& $X_0(t), \ldots, X_l(t)$ & & \\
\cline{2-2}
& $U_0(t)$ & & \\
\cline{2-2}
& $f$ & & \\
\cline{2-2}
& $S1, S2$ & & \\
\hline
\end{tabular}
\label{t1}
\end{table}
\section{Method}
\label{sec4}
We formulate a more precise vehicle dynamics for two-dimensional scenarios, propose a dimensionality-reduced set representation and provide a detailed description of the advance structure of neural network operator. The model's performance is evaluated by comparing the outputs generated by the traditional method and those produced by our model during a safety verification process, as illustrated in Figure \ref{fig1}. Additionally, multiple safety targets are utilized to quantitatively assess the model's performance.

\subsection{Characterization of More Precise Vehicle Dynamics}
We conceptualize the vehicle as an idealized bicycle model with two degrees of freedom \citep{c28}. It assumes that the vehicle operates exclusively within the x-y plane at relatively low velocities. Consequently, the vehicle dynamics can be characterized by a system comprising $n=6$ state variables (1-6) and $m=2$ control signal variables (7-8):

(1) $x_{pos}$: Vehicle's longitudinal position.

(2) $y_{pos}$: Vehicle's lateral position. 

(3) $\theta$: Vehicle's heading angle.

(4) $v$: Vehicle's longitudinal speed. 

(5) $w$: Vehicle's lateral speed. 

(6) $r$: Vehicle's yaw rate.

(7) $a_{cc}$: Vehicle's longitudinal acceleration, which is controlled by the accelerator.

(8) $\delta_f$: Vehicle's steering angle, which is controlled by the steering wheel.

The vehicle dynamics can be mathematically described by the following differential equations:

\begin{align}
    &\dot x_{pos} = v\cos(\theta)+w\sin(\theta),\\
    &\dot y_{pos} = -w\cos(\theta)+v\sin(\theta),\\
    &\dot \theta = r,\\
    &\dot v = a_{cc},\\ 
    &\dot w = -\frac{(C_{af} + C_{ar})}{m v} w - \left( \frac{b C_{ar} - a C_{af}}{m v} - v \right) r + \frac{C_{af}}{m} \delta_f, \\
    &\dot r = - \frac{b C_{ar} - a C_{af}}{I_z v} w - \frac{a^2 C_{af} + b^2 C_{ar}}{I_z v}r  + \frac{a C_{af}}{I_z} \delta_f.
\end{align}

Several parameters in the equations are intrinsic to the vehicle's physical properties. To streamline the training process, we adopt the assumption that all vehicles in our experimental scenarios possess identical parameter values. These constant parameters are specified in Table \ref{tab:vehicle_constants}.

\begin{table}[ht]
\caption{Constants for vehicle model}
\centering
\begin{tabular}{|c|c|c|c|}
\hline
\textbf{Symbol} & \textbf{Meaning} & \textbf{Value} & \textbf{Unit} \\ \hline
$m$                       & mass                       & 1500           & $kg$          \\ \hline
$I_z$                     & \makecell{moment of\\ inertia about the z axis} & 2800           & $kg \cdot m^2$ \\ \hline
$a$                       & \makecell{distance between front wheel\\ and center of mass} & 1.2            & $m$           \\ \hline
$b$                       & \makecell{distance between back wheel\\ and center of mass}   & 1.4            & $m$           \\ \hline
$C_{af}$                  & \makecell{front wheel  cornering stiffness} & 1.7$\times 10^5$        & $N / rad$ \\ \hline
$C_{ar}$                  & \makecell{back wheel cornering stiffness}  & 1.3$\times 10^5$        & $N / rad$ \\ \hline
\end{tabular}

\label{tab:vehicle_constants}
\end{table}

The idealized bicycle model, being a lower-dimensional system compared to other models, facilitates the operator's ability to discern hidden relationships between different dimensions in the data, thereby enhancing its performance. This model is also more precise than kinematic models that are independent of vehicle constants.

\subsection{Dimensionality-Reduced Set Representation}
We utilize Zonotopes to linearly characterize the relationship between the numerical values of state-space variables within a high-dimensional vector space:

Definition 1 (Zonotope): Given a center vector $c\in R^n$ and generator vectors $v_1,v_2,...,v_n\in R^n$, a Zonotope $Z\subseteq R^n$ is defined as:
\begin{equation}
    Z:=\{c+\sum^n_{i=1} \alpha_i v_i | \alpha_i\in [-1, 1]\}.\label{eq2}
\end{equation}

In a Zonotope, the generator coefficients can independently assume values within the interval, so the generator coefficient can independently take the value in $[-1, 1]$. The linear independence of the generators $v_1,v_2,...,v_n$ is typically ensured by the independence between distinct physical quantities in the state variables. Consequently, the value range for each dimension can be precisely determined through linear transformation, exhibiting symmetry about the central vector:
\begin{equation}
Z:=\{[x_1,...,x_n] | c_i-r_i\le x_i\le c_i+r_i, \forall i\in [1, n]\},\label{eq3}
\end{equation}
of which
$lb_i=c_i-r_i, ub_i=c_i+r_i.\label{eq4}$

During the traditional reachability analysis process, the Star set \citep{c19} is employed to characterize constraints deduced from the dynamic relationships among physical quantities represented by distinct dimensions of the state variable.

Definition 2 (Star): Given a center vector $c\in R^n$ and generator vectors $v_1,v_2,...,v_n\in R^n$, constraint matrix $C\in R^{p\times n}$ and constraint vector $d\in R^p$, a Star set $Star\subseteq R^n$ is defined as:
\begin{equation}
    Star:=\{c+\sum^n_{i=1} \alpha_i v_i | C\bm{\alpha} \leq d, \bm{\alpha} =[\alpha_1,\alpha_2,...,\alpha_n]\}.\label{eq5}
\end{equation}

A linear programming solver can be utilized to transform a Star set into a Zonotope and determine the range of each dimension.

It is worth noting that the feature information dimension of Zonotope is $2n$, which is on the order of $10^1$ in the vehicle system, whereas the feature information dimension of Star is on the order of $10^5$. By utilizing a set representation based on Zonotope rather than Star, we achieve a significant reduction in data dimensionality, enabling our method to perform effectively in complex vehicle systems.

\subsection{Structure of  Neural Network Operator}

Given the deduction interval $\Delta t$ and the total step $N$, the reachable set of the vehicle state at the next step is entirely determined by the reachable set of the vehicle state and control signal at the current step. Consequently, the calculation of the reachable set can be viewed as a mapping from $R^{m+n}$ to $R^{n}$, which justifies its suitability for approximation by a neural network operator. Define the reachable set of the vehicle state at the next step as $X_0^{'}\subseteq R^n$, and the reachable set of the vehicle state and the control signal at this step as $X_{U}\subseteq R^{m+n}$. The operator $F$ is defined as:
\begin{align}
    F: R^{m+n} \to R^n, X_0^{'} = F(X_{U}).
\end{align}

\begin{figure}[thpb]
    \centering
    \includegraphics[width=3.3in]{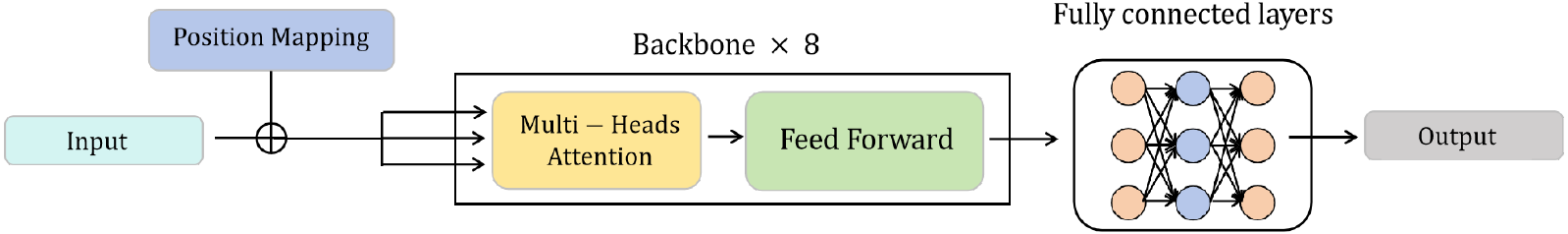} 
    \caption{Schematic diagram of neural network operator } 
    \label{fig2}
\end{figure}

Our proposed neural network operator structure is based on the Bidirectional Encoder Representations from Transformers (BERT) \citep{c22}, whose self-attention mechanism provides enhanced capability for contextual information integration and interpretation. The detailed structural configuration is illustrated in Figure \ref{fig2}. BERT is a neural network structure that can identify the nonlinear relationships between different dimensions, making it more suitable for the task of dynamic deduction compared to the network used in previous work \citep{c31}.

Different layers from input to output are as follows:

(1)Input:
The input to the operator is derived from the Zonotopes $X(t), U(t)$, corresponding to the physical variables constituting the vehicle's state and control signals, and both the center and the radius of each variable.

(2)Position mapping layer:
We choose this layer to receive input in the form of scope vectors and output higher-dimensional data.

(3)Backbone:
This layer receives input from the position mapping layer and functions as the encoder. The architecture comprises an input/output linear layer alongside $8$ blocks designed in a BERT-style configuration. Each block integrates a multi-head self-attention mechanism with a feed-forward neural network (as illustrated in Figure \ref{fig2}). 

(4)Fully connected layers:
These layers receive input from the backbone and function as the decoder. They consist of two fully connected layers preceding the final output. 

(5)Activation function:
We employ the Gaussian Error Linear Unit (GELU) \citep{c25} as the activation function following the backbone and the first fully connected layer.

(6)Output:
Ultimately, the output vector is transformed into a Zonotope $X(t+\Delta t)$.

\subsection{Data Generation and Training Settings  }
To train the neural network operator $F$, we employ traditional reachability analysis approach \citep{c18} to generate both the training data and corresponding labels for our model. We construct a specific scenario on traditional reachability analysis platforms \citep{c23} by providing the dynamic functions of the vehicles. Each vehicle's state is deduced using a dedicated platform object, which accepts $\Delta t, N$, and the vehicle's $X, U$ in the form of Zonotopes. In each step, the platform object transforms the set into a Star (Eq. \eqref{eq5}), encapsulating the deduction result in the change of the center vector and constraint matrix, and then reconverts the Star back into a Zonotope for output. After each step, the platform objects update their states to the controller to obtain new control signals or retain the existing control signals unchanged.

To collect data and labels for training, we initialize every variable in the vehicle system uniformly. For each initial value, we run the platforms for $N$ steps. In each step, for each platform object, we gather the Zonotopes $X(t), U(t)$ as the data and the Zonotope $X(t+\Delta t)$ as the label. Subsequently, both the data and the label are converted into the form of scope vectors using Eq. (\ref{eq3}-\ref{eq4}) to align with the model's input and output format.

The data generation process ensures that the operator can capture the operational principles of traditional reachability analysis, achieving comparable accuracy to conventional methods while significantly reducing computational time. Note that we generate data offline to train neural network operators. Therefore, the primary complexity is shifted to the offline training process, enabling the online application.

\begin{figure}[thpb]
    \centering
    \includegraphics[width=2.5in]{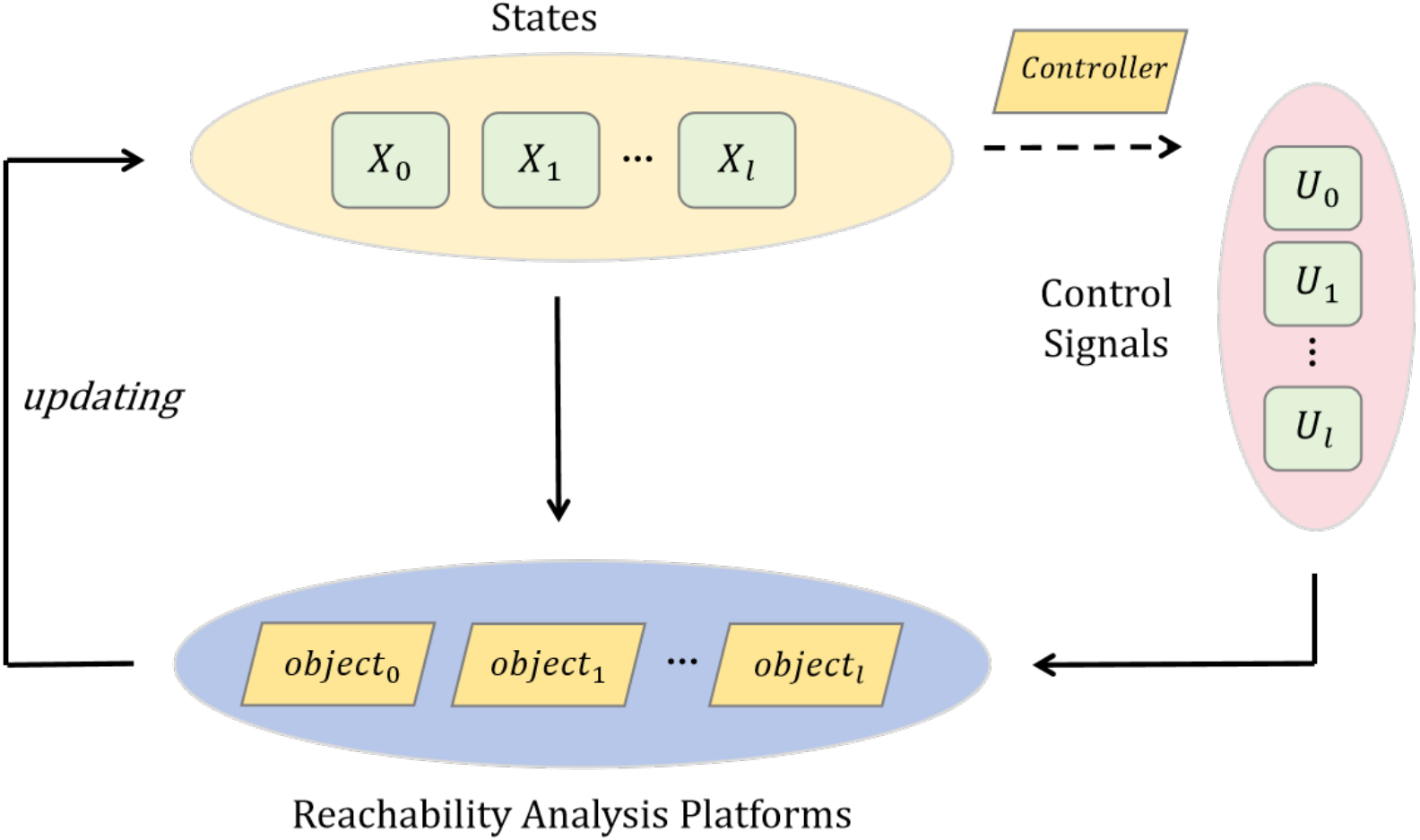} 
    \caption{An overview of deduction process of traditional method} 
    \label{fig3}
\end{figure}

The L1 Loss is used as the loss function for training the model. We utilize RMSprop \citep{c26} as the optimizer to perform stochastic gradient descent and MultiStepLR \citep{c27} as the learning rate scheduler for managing the training process. The scheduler is configured to multiply the learning rate by 0.5 every 150 epochs.

\subsection{Safety Target}
As the operator's safety and effectiveness can't be directly indicated by loss, we design two safety targets $recall$ and $precision$ to evaluate them. 

$recall$ describes the coverage ratio of the reachable set represented by operator's output to reachable set represented by label. Larger $recall$ means higher accuracy. This target is selected to ensure that the reachable set computed by the operator sufficiently encompasses the reachable set represented by the label. It is defined as: $recall = \frac{|R_{model}\cap R_{label}|}{|R_{label}|} \in [0, 1],$

of which $|R|$ means the 'volume' of the reachable set in high dimensional space.

For complex systems involving more than three variables, evaluating the recall on the affine set of combinations of multiple variables becomes more meaningful. The following example will be utilized in our experiment.

$recall_{pos}$ represents the recall on vehicle's position in two dimensional scenarios, defined as:
\begin{align}
recall_{pos} &= \frac{|R_{m.,(x,y)}\cap R_{l.,(x,y)}|}{|R_{l.,(x,y)}|} \in [0, 1] \\
&= \frac{cover_x\cdot cover_y}{(ub_{l.,x}-lb_{l.,x})(ub_{l.,y}-lb_{l.,y})}, 
\end{align}
in which :
\begin{align}
cover_{x} &= min(ub_{l.,x},ub_{m.,x})-max(lb_{l.,x},lb_{m.,x}),\\
cover_{y} &= min(ub_{l.,y},ub_{m.,y})-max(lb_{l.,y},lb_{m.,y}).
\end{align}

In the equations above, $l.$ is for $label$, $m.$ is for $model$, and $x,y$ mean the two dimensions in vehicle's position.

$precision$ describes the coverage ratio of the reachable set represented by label to reachable set represented by model output. Larger $precision$ means lower model conservativeness. We select this target to assess whether the operator erroneously predicts numerous unreachable states as reachable. It is defined as $precision = \frac{|R_{model}\cap R_{label}|}{|R_{model}|} \in [0, 1]$. Similar to $recall_{pos}$, we can define $precision_{pos}$.

These two targets offer a representative evaluation of the accuracy and conservativeness of the neural network operator. However, these targets are not integrated into the training process of the operators due to the lack of suitable implementation mechanisms. Therefore, a potential limitation of the neural network operator is that the $recall$ and the $precision$ may not achieve a value of 1, which could lead to a small probability of inaccuracies in safety verification.

\section{Experiments}
\label{sec5}
We train the model and design several typical driving scenarios, conduct the safety verification process and compare our operator's performance against classical methods.

\subsection{Model Training}
We collect the data using the described approach and divide it into training, test, and validation sets in a 5:1:1 ratio. The training set contains 1,000,000 samples, while both the test and validation sets comprise 200,000 samples each. The model is trained for 600 epochs, utilizing one NVIDIA GeForce RTX 3090 GPU. The final loss and the model's $recall_{pos}$ and $precision_{pos}$ for one step, averaged across the entire dataset, are listed as in Table \ref{tab:performance_metrics}.

\begin{table}[ht]
\caption{Neural network operator's performance on different sets}
\centering
\begin{tabular}{|c|c|c|c|}
\hline
Datasets & $loss$ & $recall_{pos}$ & $precision_{pos}$ \\ \hline
Train & $0.0021$ & $0.9965 \pm 0.0003$ & $0.9953 \pm 0.0003$ \\ \hline
Test & $0.0026$ & $0.9926 \pm 0.0006$ & $0.9905 \pm 0.0005$ \\ \hline
Validation & $0.0027$ & $0.9918 \pm 0.0006$ & $0.9904 \pm 0.0006$ \\ \hline
\end{tabular}

\label{tab:performance_metrics}
\end{table}

\subsection{Vehicle Boundary Representation}
\begin{figure}[thpb]
    \centering
    \includegraphics[width=1.7in]{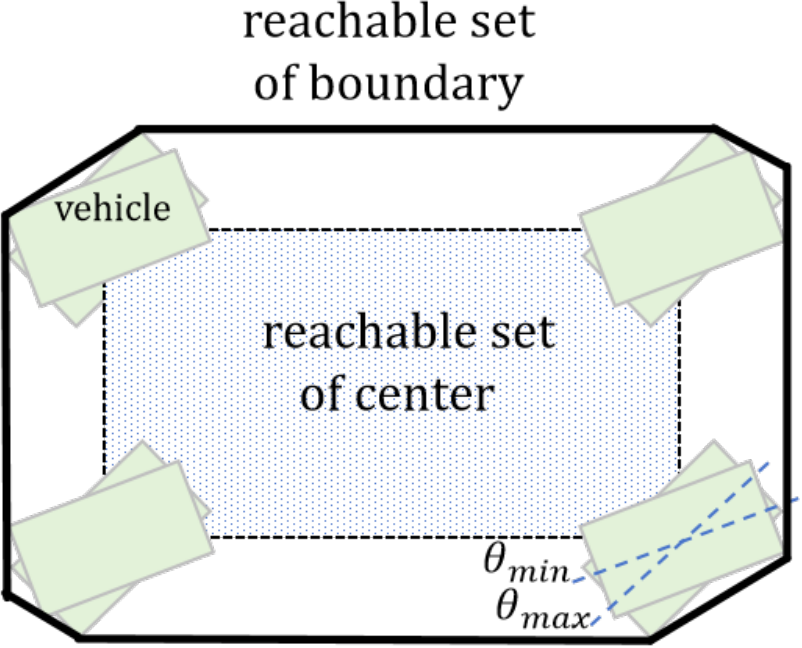} 
    \caption{Vehicle boundary octagon} 
    \label{fig5}
\end{figure}

We define the vehicle's shape as a rectangle measuring $3.5m$ in length and $1.8m$ in width. Given that $x_{pos}$ and $y_{pos}$ denote the position of the vehicle's geometric center and $\theta$ determines its orientation, the reachable set of the vehicle's actual boundary in the x-y plane can be represented by an octagon, as shown in Figure \ref{fig5}. Octagons will be used for safety assessments below.

\subsection{Single Vehicle Scenario}
\label{SVS}

This scenario involves a single autonomous vehicle (AV) driving along a straight road in the positive direction of the X-axis. For each experiment in this scenario, initial center values for physical variables are randomly selected within a specified range. A random control signal is also applied and remains constant throughout the verification process. To account for uncertainty, each physical variable is assigned a small radius, forming a range variable with the randomly selected center value.

The X-axis is centered on a five-meter-wide road, assumed to be of sufficient length. The vehicle's behavior must adhere to an $S1$ standard, ensuring that the vehicle's boundary does not exceed $2.5m$ or fall below $-2.5m$ at each step. The initial values used in this scenario are listed in Table \ref{table5}.

\begin{table}[ht]
\centering
\caption{the initial values of AV's physical variables}
\begin{tabular}{|c|c|c|c|}
\hline
\textbf{Variables} & \textbf{Center Value} & \textbf{Radius Range} & \textbf{Unit} \\
\hline
$x_{pos}$ & $[1, 15]$ & $(0.5,1)$ & $m$ \\
\hline
$y_{pos}$ & $[-2, 2]$ & $(0.25,0.5)$ & $m$ \\
\hline
$\theta$ & $[-4, 4] \times \frac{\pi}{180}$ & $(0.25,0.5) \times \frac{\pi}{180}$ & $rad$ \\
\hline
$v$ & $[4, 8]$ & $(0.25,0.5)$ & $m/s$ \\
\hline
$w$ & $[-0.6, 0.6]$ & $(0.05,0.1)$ & $m/s$ \\
\hline
$r$ & $[-0.1, 0.1]$ & $(0.01,0.02)$ & $rad/s$ \\
\hline
$a_{cc}$ & $[-3, 3]$ & $(0.1,0.2)$ & $m/s^2$ \\
\hline
$\delta_f$ & $[-5, 5] \times \frac{\pi}{180}$ & $(0.05,0.1) \times \frac{\pi}{180}$ & $rad$ \\
\hline
\end{tabular}

\label{table5}
\end{table}

We randomly generate $100$ experiments in this scenario and compare the results obtained from the traditional method (NNV) with those from our operator (NN Operator). The evaluation targets include the deduction time for one step, $recall_{pos}$ and $precision_{pos}$ for each step, and the accuracy rate of the operator's safety judgment. Although the results of the NN Operator and NNV are not entirely consistent in a single step, this discrepancy does not affect the safety judgment accuracy in these cases. Across these 100 experiments, the safety judgments given by the NN Operator demonstrated complete consistency with those given by NNV. The remaining results are summarized in Table \ref{table6}. It is worth noting that the $recall$ and $precision$ are calculated using the results of NNV as reference values, and thus these two targets for NNV are naturally 1.

\begin{table}[ht]
\centering
\caption{AV's results in single vehicle scenario}
\label{your_label_here}
\renewcommand{\arraystretch}{1.5}
\begin{tabular}{|c|c|c|c|}
\hline
\multicolumn{2}{|c|}{\textbf{Targets}} & \textbf{NNV} & \textbf{NN Operator} \\
\hline
\multirow{5}{*}{$recall_{pos}$} 
& step 1 & $1$ & $0.9855 \pm 0.0141$ \\
\cline{2-4}
& step 2 & $1$ & $0.9790 \pm 0.0132$ \\
\cline{2-4}
& step 3 & $1$ & $0.9701 \pm 0.0168$ \\
\cline{2-4}
& step 4 & $1$ & $0.9602 \pm 0.0213$ \\
\cline{2-4}
& step 5 & $1$ & $0.9508 \pm 0.0258$ \\
\hline
\multirow{5}{*}{$precision_{pos}$} 
& step 1 & $1$ & $0.9886 \pm 0.0159$ \\
\cline{2-4}
& step 2 & $1$ & $0.9902 \pm 0.0129$ \\
\cline{2-4}
& step 3 & $1$ & $0.9895 \pm 0.0136$ \\
\cline{2-4}
& step 4 & $1$ & $0.9874 \pm 0.0160$ \\
\cline{2-4}
& step 5 & $1$ & $0.9848 \pm 0.0192$ \\
\hline
\multicolumn{2}{|c|}{time per step (s)} & $10.2513 \pm 0.4473$ & $0.0990 \pm 0.0028$ \\
\hline
\end{tabular}

\label{table6}
\end{table}

\begin{figure}[htb]
	\centering
	\begin{minipage}{0.49\linewidth}
		\vspace{3pt}
		\centerline{\includegraphics[width=\textwidth]{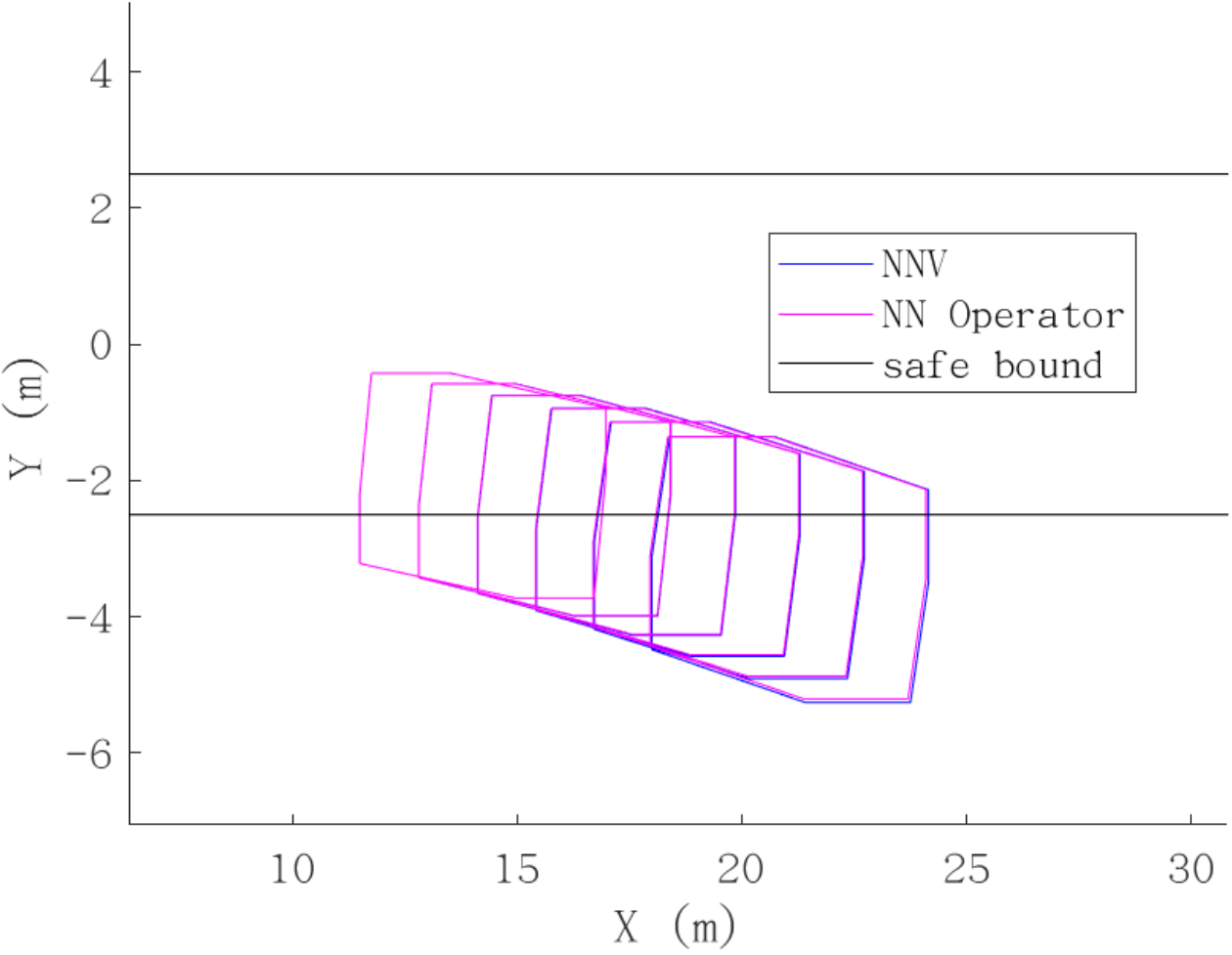}}
		\centerline{Demo I}
	\end{minipage}
	\begin{minipage}{0.49\linewidth}
		\vspace{3pt}
		\centerline{\includegraphics[width=\textwidth]{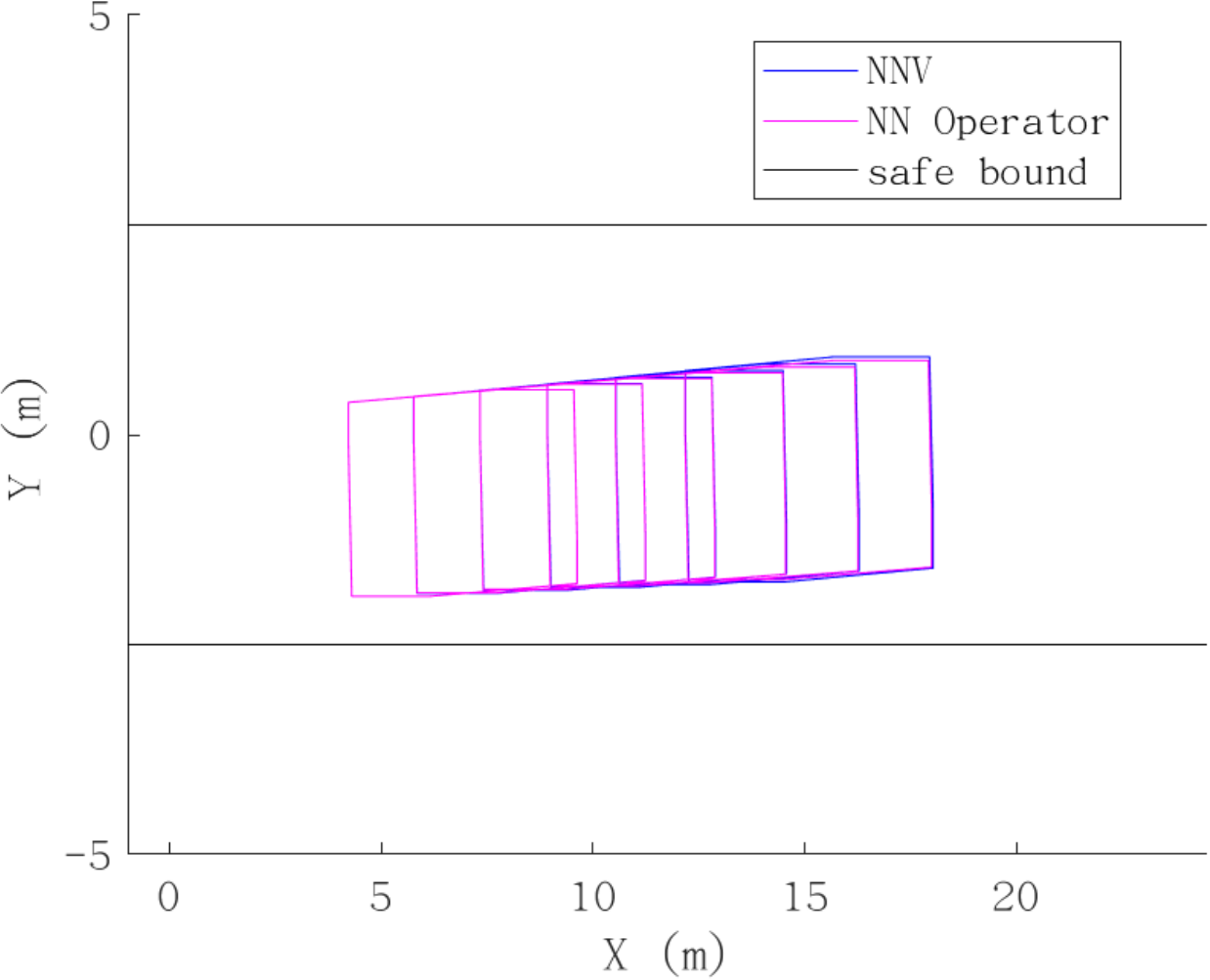}}
		\centerline{Demo II}
	\end{minipage}
 
	\caption{Demos for Single Vehicle Scenario}
        \label{fig4}
\end{figure}

We present two demos of the single vehicle scenario in Figure \ref{fig4}. The left demo and the right demo respectively illustrates an unsafe example and a safe one. In these figures, the two black lines indicating the safe bounds at $y=2.5$ and $y=-2.5$. The blue and the pink octagons respectively depict the reachable sets calculated by the traditional method (NNV) and our NN Operator. As demonstrated, the NN Operator accurately judges safety, even when the vehicle is close to the safety bounds.

\subsection{Double Vehicles Scenario}
\label{DVS}

This scenario involves one AV and one background vehicle (BV) driving in the positive direction of the X-axis. The BV is positioned in front of the AV with a randomly close distance. Other variables are sampled independently. Notably, the control signals for the BVs are selected to cover the maximum range that the center value can achieve. The physical variables for the AV and BV are consistent with those outlined in the single vehicle scenario (see Table \ref{table5}), excluding the position variables and the control signals for BVs. For safety judgment, the AV must comply with the $S1$ and $S2$ standards that prevent the reachable sets of the AV and BV from overlapping.

We randomly generate $100$ experiments for this scenario. Across these experiments, the safety judgments given by the NN Operator demonstrated complete consistency with those given by NNV. The targets for the background vehicle are provided in Table \ref{table8}.

\begin{table}[ht]
\centering
\caption{BV's results in double vehicle scenario}
\label{table8}
\renewcommand{\arraystretch}{1.5}
\begin{tabular}{|c|c|c|c|}
\hline
\multicolumn{2}{|c|}{\textbf{Targets}} & \textbf{NNV} & \textbf{NN Operator} \\
\hline
\multirow{5}{*}{$recall_{pos}$} 
& step 1 & $1$ & $0.9925 \pm 0.0071$ \\
\cline{2-4}
& step 2 & $1$ & $0.9843 \pm 0.0099$ \\
\cline{2-4}
& step 3 & $1$ & $0.9690 \pm 0.0122$ \\
\cline{2-4}
& step 4 & $1$ & $0.9762 \pm 0.0126$ \\
\cline{2-4}
& step 5 & $1$ & $0.9719 \pm 0.0152$ \\
\hline
\multirow{5}{*}{$precision_{pos}$} 
& step 1 & $1$ & $0.9872 \pm 0.0113$ \\
\cline{2-4}
& step 2 & $1$ & $0.9892 \pm 0.0105$ \\
\cline{2-4}
& step 3 & $1$ & $0.9880 \pm 0.0106$ \\
\cline{2-4}
& step 4 & $1$ & $0.9813 \pm 0.0128$ \\
\cline{2-4}
& step 5 & $1$ & $0.9739 \pm 0.0178$ \\
\hline
\multicolumn{2}{|c|}{time per step (s)} & $11.6613 \pm 0.2746$ & $0.0836 \pm 0.0087$ \\
\hline
\end{tabular}
\end{table}

\begin{figure}[htb]

	\begin{minipage}{0.45\linewidth}
		\vspace{1pt}
		\centerline{\includegraphics[width=\textwidth]{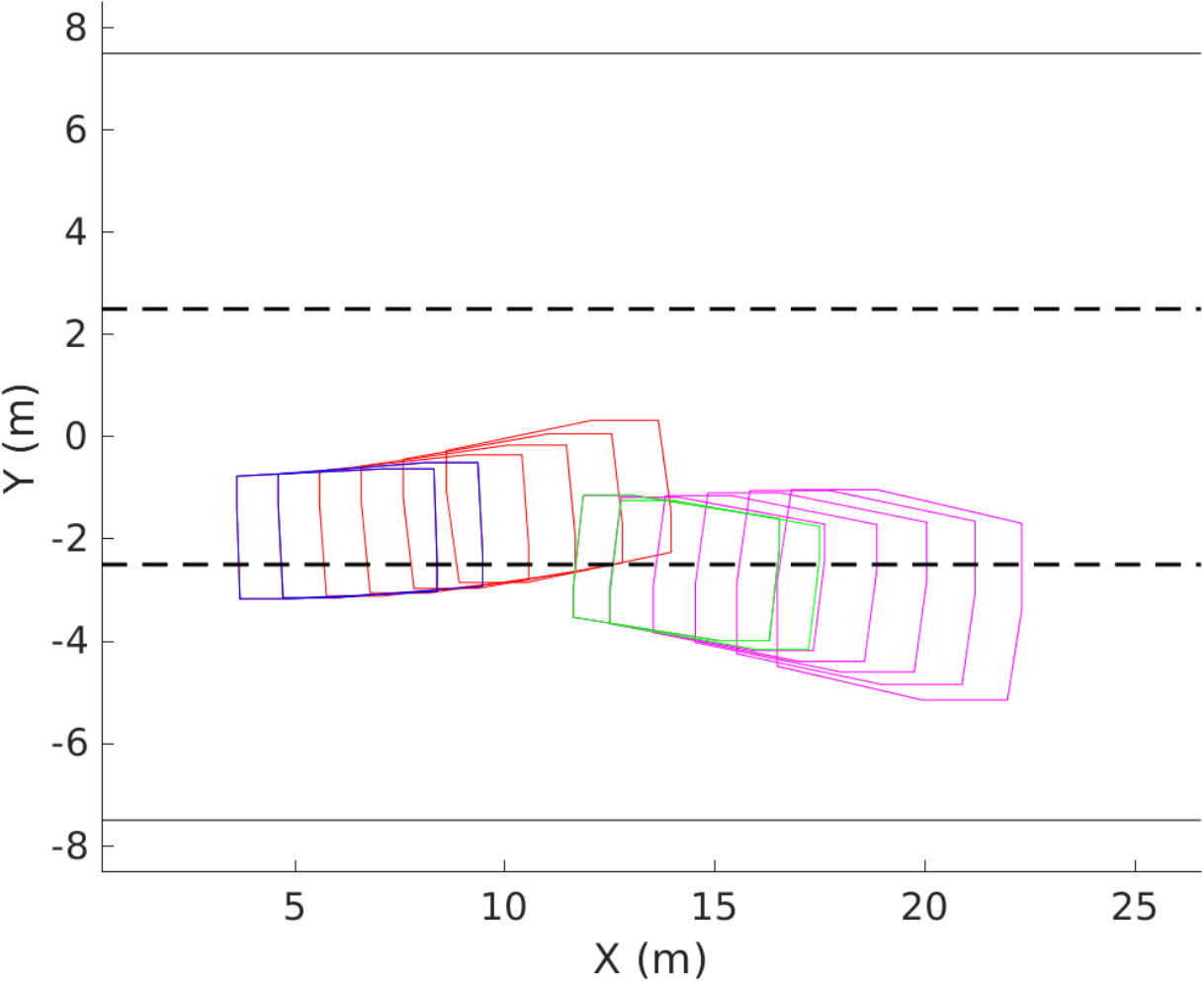}}
		\centerline{Demo I}
	\end{minipage}
        \begin{minipage}{0.45\linewidth}
		\vspace{1pt}
		\centerline{\includegraphics[width=\textwidth]{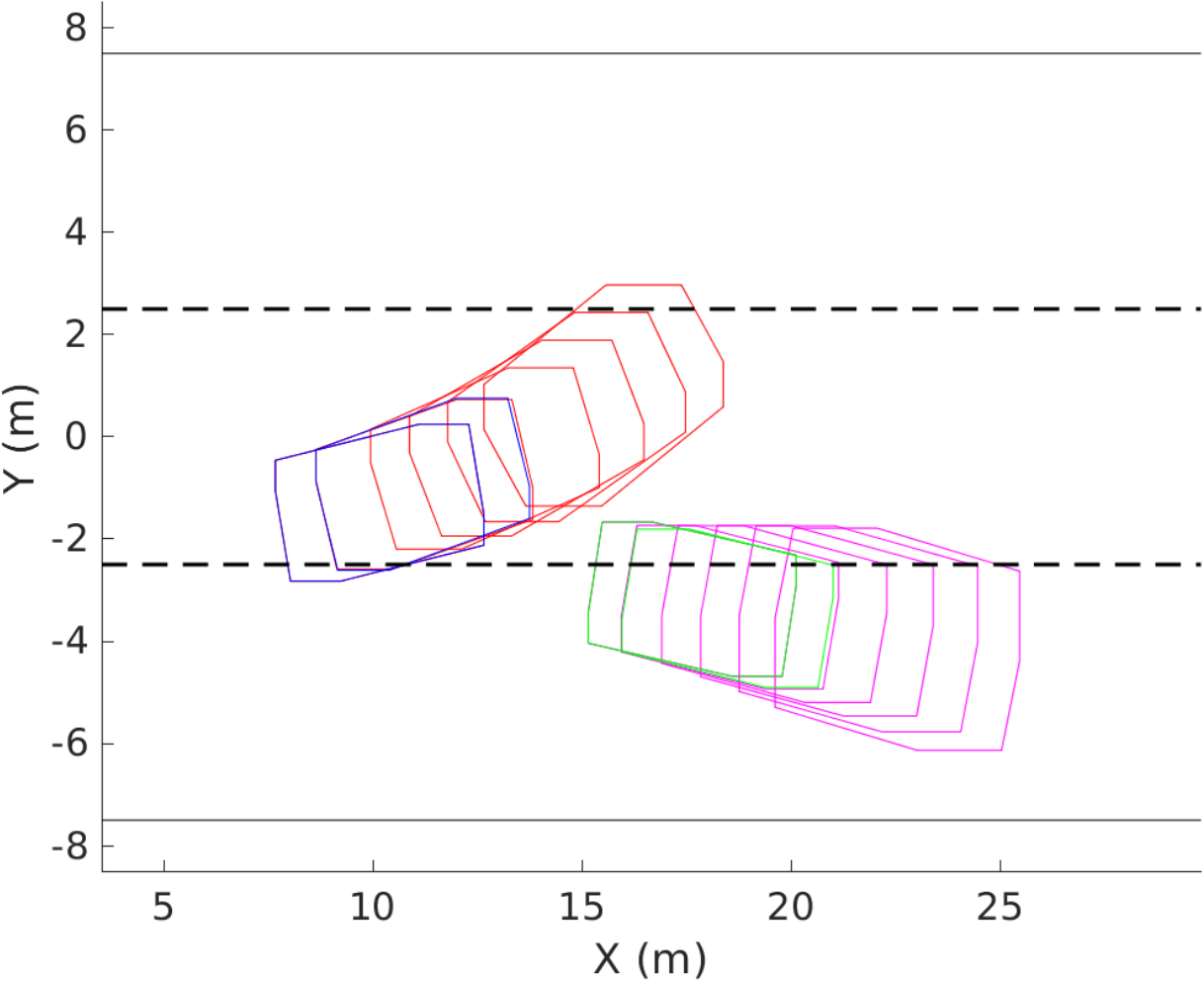}}
		\centerline{Demo II}
	\end{minipage}
 
	\caption{Demos for Double Vehicle Scenario}
        \label{fig6}
\end{figure}

We illustrate different slices of a double-vehicle scenario with real-time safety verification in Figure \ref{fig6}. In this scenario, each step of the AV requires a complete safety verification process lasting $1s$. During the process, the initial control signal of the AV is verified by the NN Operator for $1s$. If the signal is safe, it is accepted. If unsafe, the AV executes a sudden brake (with $acc = -3, deltaf=0$) and terminates the safety verification process. The state of AV after $0.2s$ is updated by the NNV. Subsequently, a new initial control signal is generated by AV controller. The AV's control signal is given by the Intelligent Driver Model (IDM) \citep{c29} and the MOBIL model \citep{c30}. The BV is controlled based on the linear feedback of $v$ and $y_{pos}$.

In demonstrations, the lines denote a three-lane road   . The blue and green octagons respectively depict the reachable set of the AV's and BV's boundary before and after each step. Red and pink octagons indicate the predicted boundaries of the AV and BV. In Figure \ref{fig6}, the AV completes an upward lane change with BV ahead.

\vspace{-7pt}
\subsection{Multi Vehicles Scenario}

In this scenario, one AV and five BVs are driving on a straight road aligned with the positive X-axis. The scenario settings, safety standards, and the initial values of physical variables are identical to those in above scenarios. 

\begin{figure}[thpb]
    \centering
    \includegraphics[width=3.5in]{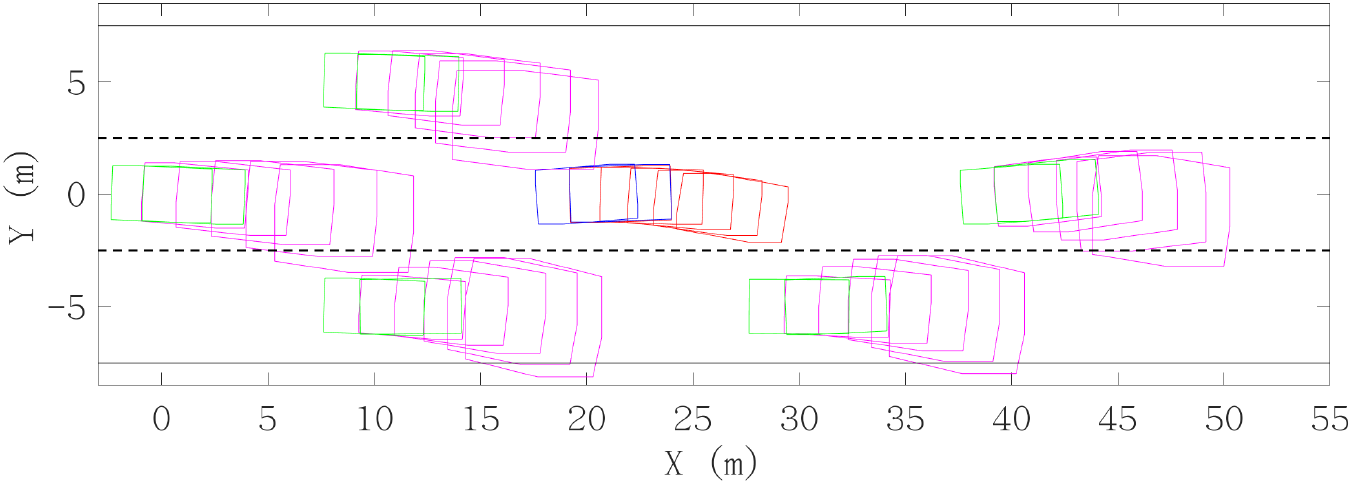} 
    \caption{Demos for Multi Vehicle Scenario} 
    \label{fig8}
\end{figure}

Figure \ref{fig8} illustrates a slice representing a frame of a continuous scenario where the AV attempts to change lanes rightward while surrounded by five BVs. 

\vspace{-5pt}
\section{Conclusions}
\label{sec6}
In conclusion, our safety verification method demonstrates a significant improvement in efficiency, being over $100$ times faster than traditional reachability analysis methods, without substantially compromising accuracy or increasing conservativeness. This efficiency enables real-time safety verification for autonomous vehicles in general scenarios, marking a critical advancement in their operational safety. Our method also establishes a novel safety verification framework, enabling the neural network operator to handle scalable vehicle systems effectively.

The primary limitation of our method lies in the neural network operator's current capability to handle only specific vehicle dynamics and fixed vehicle constants. This limitation is also inherent in classical methods. We expect that incorporating the constants as an additional input during the generation of derivation data for different vehicles and integrating it into the training process of the neural network operator could enhance its performance. 

\bibliography{ifacconf}            
\end{document}